\documentclass[aps,prl,onecolumn,superscriptaddress]{revtex4}

\usepackage{graphicx}
\usepackage{amsmath}
\usepackage{amsfonts}
\usepackage{bm}
\usepackage{color}

\newcommand{\ket}[1]{\left\vert#1\right\rangle}
\newcommand{\bra}[1]{\langle #1\vert}

\begin{document}
\title{Quantum teleportation using active feed-forward between two Canary Islands}
\author{Xiao-song Ma}
\affiliation{Institute for Quantum Optics and Quantum Information (IQOQI), Austrian
Academy of Sciences, Boltzmanngasse 3, A-1090 Vienna, Austria}
\affiliation{Vienna Center for Quantum Science and Technology, Faculty of Physics, University of Vienna, Boltzmanngasse 5, A-1090 Vienna,
Austria}
\author{Thomas Herbst}
\affiliation{Faculty of Physics, University of Vienna, Boltzmanngasse 5, A-1090 Vienna,
Austria}
\author{Thomas Scheidl}
\affiliation{Institute for Quantum Optics and Quantum Information (IQOQI), Austrian
Academy of Sciences, Boltzmanngasse 3, A-1090 Vienna, Austria}
\author{Daqing Wang}
\affiliation{Institute for Quantum Optics and Quantum Information (IQOQI), Austrian
Academy of Sciences, Boltzmanngasse 3, A-1090 Vienna, Austria}
\author{Sebastian Kropatschek}
\affiliation{Institute for Quantum Optics and Quantum Information (IQOQI), Austrian
Academy of Sciences, Boltzmanngasse 3, A-1090 Vienna, Austria}
\author{William Naylor}
\affiliation{Institute for Quantum Optics and Quantum Information (IQOQI), Austrian
Academy of Sciences, Boltzmanngasse 3, A-1090 Vienna, Austria}
\author{Alexandra Mech}
\affiliation{Vienna Center for Quantum Science and Technology, Faculty of Physics, University of Vienna, Boltzmanngasse 5, A-1090 Vienna,
Austria}
\affiliation{Institute for Quantum Optics and Quantum Information (IQOQI), Austrian
Academy of Sciences, Boltzmanngasse 3, A-1090 Vienna, Austria}
\author{Bernhard Wittmann}
\affiliation{Vienna Center for Quantum Science and Technology, Faculty of Physics, University of Vienna, Boltzmanngasse 5, A-1090 Vienna,
Austria}
\affiliation{Institute for Quantum Optics and Quantum Information (IQOQI), Austrian
Academy of Sciences, Boltzmanngasse 3, A-1090 Vienna, Austria}
\author{Johannes Kofler}
\affiliation{Max Planck Institute of Quantum Optics, Hans-Kopfermann-Str.\ 1, 85748 Garching/Munich, Germany}
\affiliation{Institute for Quantum Optics and Quantum Information (IQOQI), Austrian
Academy of Sciences, Boltzmanngasse 3, A-1090 Vienna, Austria}
\author{Elena Anisimova}
\affiliation{Institute for Quantum Computing and Department of Physics and Astronomy, University of Waterloo, 200 University Avenue West, Waterloo, ON, N2L 3G1, Canada}
\affiliation{Department of Electronics and Telecommunications, Norwegian University of Science and Technology, NO-7491 Trondheim, Norway}
\author{Vadim Makarov}
\affiliation{Institute for Quantum Computing and Department of Physics and Astronomy, University of Waterloo, 200 University Avenue West, Waterloo, ON, N2L 3G1, Canada}
\affiliation{Department of Electronics and Telecommunications, Norwegian University of Science and Technology, NO-7491 Trondheim, Norway}
\author{Thomas Jennewein}
\affiliation{Institute for Quantum Computing and Department of Physics and Astronomy, University of Waterloo, 200 University Avenue West, Waterloo, ON, N2L 3G1, Canada}
\author{Rupert Ursin}
\affiliation{Institute for Quantum Optics and Quantum Information (IQOQI), Austrian
Academy of Sciences, Boltzmanngasse 3, A-1090 Vienna, Austria}
\author{Anton Zeilinger}
\affiliation{Institute for Quantum Optics and Quantum Information (IQOQI), Austrian
Academy of Sciences, Boltzmanngasse 3, A-1090 Vienna, Austria}
\affiliation{Vienna Center for Quantum Science and Technology, Faculty of Physics, University of Vienna, Boltzmanngasse 5, A-1090 Vienna, Austria}
\affiliation{Faculty of Physics, University of Vienna, Boltzmanngasse 5, A-1090 Vienna, Austria}

\date{\today}

\begin{abstract}
Quantum teleportation~\cite{Bennett1993} is a quintessential prerequisite of many quantum information processing protocols~\cite{NielsenChuang2000,Gottesmann1999,Knill2001}. By using quantum teleportation, one can circumvent the no-cloning theorem~\cite{Wootters1982} and faithfully transfer unknown quantum states to a party whose location is even unknown over arbitrary distances. Ever since the first experimental demonstrations of quantum teleportation of independent qubits~\cite{Bouwmeester1997} and of squeezed states~\cite{Furusawa1998}, researchers have progressively extended the communication distance in teleportation, usually without active feed-forward of the classical Bell-state measurement result which is an essential ingredient in future applications such as communication between quantum computers. Here we report the first long-distance quantum teleportation experiment with active feed-forward in real time. The experiment employed two optical links, quantum and classical, over 143 km free space between the two Canary Islands of La Palma and Tenerife. To achieve this, the experiment had to employ novel techniques such as a frequency-uncorrelated polarization-entangled photon pair source, ultra-low-noise single-photon detectors, and entanglement-assisted clock synchronization. The average teleported state fidelity was well beyond the classical limit of 2/3. Furthermore, we confirmed the quality of the quantum teleportation procedure (without feed-forward) by complete quantum process tomography. Our experiment confirms the maturity and applicability of the involved technologies in real-world scenarios, and is a milestone towards future satellite-based quantum teleportation.
\end{abstract}

\maketitle


\section{Introduction}
Recently, significant progress has been made in the field of quantum communication based on optical free-space links~\cite{Hughes2002,Kurtsiefer2002,Aspelmeyer2003,Ursin2007,Villoresi2008,Fedrizzi2009,Scheidl2009,Jin2010,Scheidl2010}, which potentially allow much larger propagation distances compared to the existing fiber networks because of the lower photon loss per kilometer. In order to enable quantum communication on a global scale and among parties not having access to any fiber network, it is foreseeable that experiments will involve ground-to-satellite~\cite{Villoresi2008} and inter-satellite links.

Previous experiments focused on the distribution of quantum states of single photons or entangled photon pairs over optical free-space links. However, the realization of the more sophisticated multiphoton quantum information protocols, such as quantum teleportation, remained an experimental challenge under real-world conditions. Quantum teleportation is based on the simultaneous creation of at least three photons, which for random photon pair sources reduces the count rate by several orders of magnitude compared to experiments using only two photons~\cite{Fedrizzi2009,Scheidl2010}. This decreases the signal-to-noise ratio and requires a long integration time, such that high system stability is necessary. Moreover, the complexity and environmental requirements of a quantum teleportation setup are increased significantly compared to previous two-photon experiments, which provides significant experimental and technological challenges. Most earlier teleportation~\cite{Bouwmeester1997,Furusawa1998} and remote-state preparation variant~\cite{Boshi1998} were in-lab demonstrations, and hence the communication distances were rather limited. Although fiber-based teleportation has been demonstrated experimentally~\cite{Marcikic2003,Ursin2004}, the maximum transmission distance is limited by intrinsic photon losses in optical fiber, unless quantum repeaters are involved~\cite{Briegel1998,Duan2001}.

\begin{figure}
\includegraphics[width=0.85\textwidth]{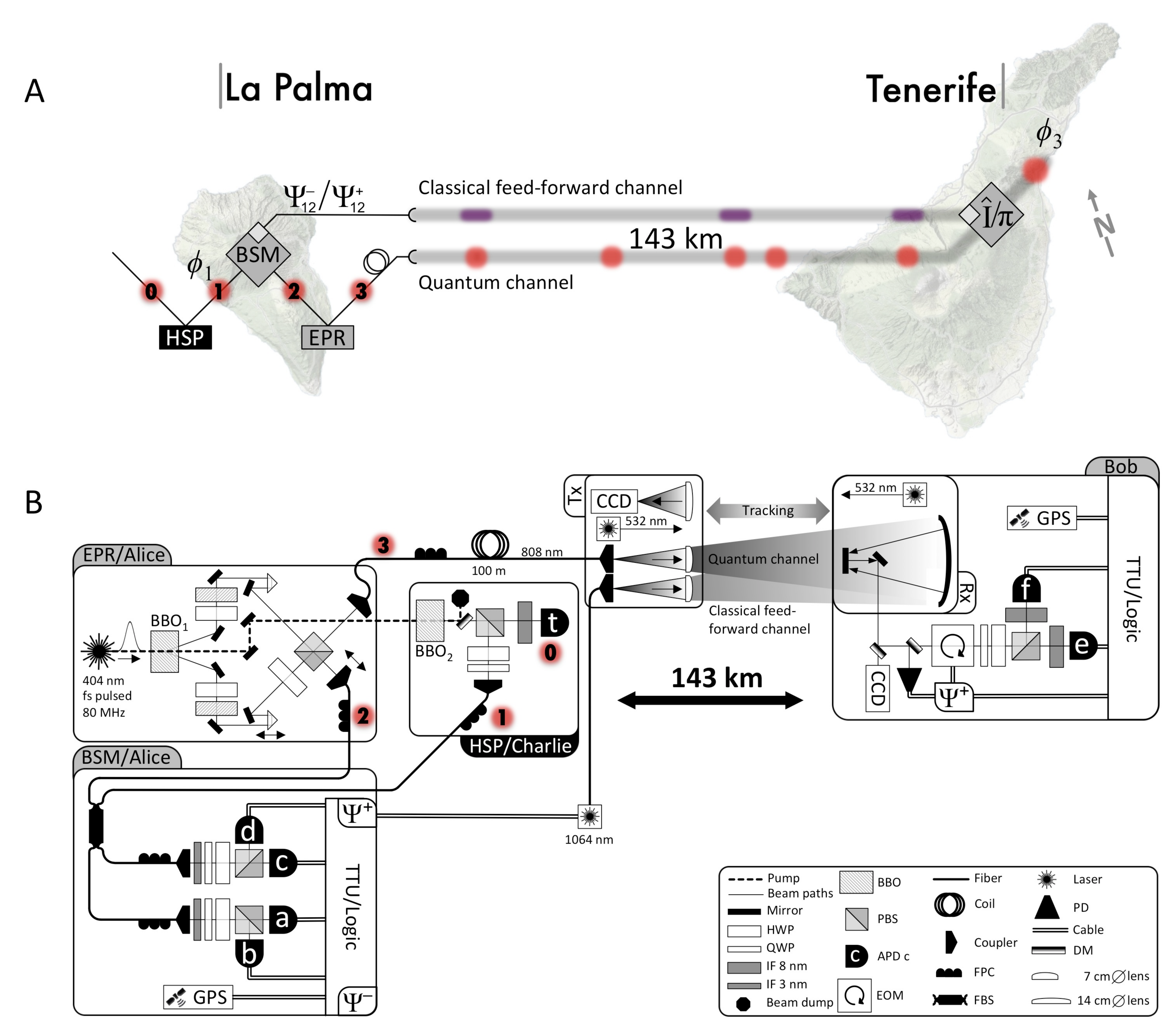}
\caption{\label{figSetup}Quantum teleportation between the Canary Islands La Palma and Tenerife over both quantum and classical 143~km free-space channels. (A) Scheme: Alice and Charlie are situated in La Palma, and Bob in Tenerife. Charlie prepares teleportation input photon 1 in $\ket{\phi}_1$, using a heralded single-photon (HSP) source with a trigger photon 0. An Einstein-Podolsky-Rosen (EPR) source generates an entangled pair of photons 2 and 3. Alice then performs a Bell-state measurement (BSM) on photons 1 and 2 and projects them onto two of the four Bell states ($\ket{\Psi^-}_{12}$/$\ket{\Psi^+}_{12}$), and sends the result via the classical feed-forward channel to Bob. Photon 3 is sent via the free-space quantum channel to Bob, who applies a unitary transformation (\^{\textrm{I}}/$\pi$) on photon 3 depending on the BSM result and thus turns its state $\ket{\phi}_3$ into a replica of the initial quantum state. (B) Setup: In La Palma, a frequency-uncorrelated polarization-entangled photon pair source generated photons 2 and 3 in BBO$_1$ (EPR/Alice) and a collinear photon pair source generated photons 0 and 1 in BBO$_2$ (HSP/Charlie). All single photons were coupled into single-mode fibers. For implementing the BSM, photons 1 and 2 interfered in a fiber beam splitter (FBS) followed by polarization-resolving single-photon detection (BSM/Alice). Photon 3 was guided to the transmitter telescope using a 100~m single-mode fiber and sent to Bob in Tenerife, where the unitary transformation was implemented using an electro-optical modulator (EOM) and its polarization was measured. A real-time feed-forward operation was implemented by encoding the $\ket{\Psi^+}_{12}$ BSM result in the 1064~nm laser pulses, which were then sent to Bob via the feed-forward channel. On Bob's side, they were detected with a photodetector (PD) and used to trigger the EOM to perform the required $\pi$ phase shift operation. See main text for details.}
\end{figure}

Quantum teleportation relies on using both a quantum channel and a classical channel between two parties, usually called Alice and Bob~\cite{Bennett1993} (shown in Fig.~\ref{figSetup}A). The quantum channel is used by Alice and Bob to share the entangled auxiliary state
\begin{equation} \label{bellstate}
|\Psi^- \rangle_{23}  =
\frac{1}{\sqrt{2}}(|H\rangle_2 |V\rangle_3 - |V\rangle_2
|H\rangle_3),
\end{equation}
which is one of the four maximally entangled Bell states ($\ket{\Psi^\pm}=\frac{1}{\sqrt{2}}(|H\rangle |V\rangle \pm |V\rangle
|H\rangle)$ and $\ket{\Phi^\pm}=\frac{1}{\sqrt{2}}(|H\rangle |H\rangle \pm |V\rangle|V\rangle)$).
$|H\rangle_j$ and $ |V\rangle_j$ denote the horizontal and
vertical polarization states of photon $j$. Alice and Bob share this entangled state, where photon 2 is with Alice and photon 3 is with Bob. Charlie provides the input photon 1 to be teleported to Alice in a general polarization state:
\begin{equation} \label{photon1}
|\phi\rangle_1=\alpha
\ket{H}_1 + \beta \ket{V}_1,
\end{equation}
where $\alpha$ and $\beta$ are complex numbers ($|\alpha|^2 +|\beta|^2=1$), unknown to both Alice and Bob.

Alice then performs a Bell-state measurement (BSM), projecting photons 1 and 2 randomly onto one of the four Bell states each with the same probability of $25\%$. As a consequence, photon 3 is projected onto the input state $|\phi\rangle_1$, up to a unitary transformation (\textbf{U}) which depends on the outcome of the BSM. When Alice feeds the outcome of the BSM forward to Bob via the classical channel, he can implement the corresponding unitary operation in real time and thus obtain photon 3 in the initial state (2) of photon 1. If $\ket{\Psi^-}_{12}$ is detected, then \textbf{U} corresponds to the identity operation which means that Bob needs not to do anything. If, on the other hand, $\ket{\Psi^+}_{12}$ is detected, Bob has to apply a $\pi$ phase shift between the horizontal and the vertical component of his photon 3.

\section{Experiment}

Our experiment was conducted between Alice's transmitter station at the Jacobus Kapteyn Telescope (JKT) of the Isaac Newton Group on La Palma and Bob's receiver station at the Optical Ground Station (OGS) of the European Space Agency on Tenerife, separated by 143~km, both at altitudes of about 2400~m. Our experimental setup is shown in Fig.~\ref{figSetup}B. At La Palma, near-infrared femtosecond pulses (central wavelength of 808~nm) emitted from a mode-locked Ti:Sapphire laser were up-converted to blue pulses (central wavelength of 404~nm), with a repetition rate of 80~MHz. They were employed to generate two photon pairs via type-II spontaneous parametric down conversion (SPDC) in two non-linear $\beta$-barium borate (BBO) crystals placed in sequence. The first SPDC source was aligned to emit the entangled auxiliary photon pairs (photons 2 and 3) in the $\ket{\Psi^-}_{23}$ state~\cite{Kwiat1995}, Eq. (1), while the second crystal was a heralded single-photon (HSP) source providing Charlie's photon 1 to be teleported. That source delivered pairs of horizontally (photon 0) and vertically (photon 1) polarized photons in a product state. The detection of photon 0 by detector \textbf{t} served as a trigger to herald the presence of photon 1. All photons were spectrally filtered using interference filters (IF) and coupled into single-mode fibers for spectral and spatial mode selection. For realizing the BSM, Alice's photon 2 was overlapped on a fiber beam splitter (FBS) with the teleportation input photon 1, whose polarization was arbitrarily prepared by Charlie using half- and quarter-wave plates. In each output port of the FBS, a polarizing beam splitter allowed to project these photons on either horizontal or vertical polarization. Fiber polarization controllers (FPC) were used to compensate the unwanted polarization rotation induced by the fibers.

When photons 1 and 2 leave the FBS in different (or same) outputs and exhibit orthogonal polarization, Alice projects photons 1 and 2 onto $\ket{\Psi^-}_{12}$ (or $\ket{\Psi^+}_{12}$). All photons were detected using single-photon silicon avalanche photodiodes (Si-APDs). Note that the BSM was based on a 3-fold detection event between photons 0, 1 and 2 originating from one and the same pump pulse. This implementation of the BSM allowed us to identify $\ket{\Psi^-}_{12}$ (i.e., three-fold coincidence between detectors \textbf{t}-\textbf{a}-\textbf{d} or \textbf{t}-\textbf{b}-\textbf{c}) and $\ket{\Psi^+}_{12}$ (i.e., three-fold coincidence between detectors \textbf{t}-\textbf{a}-\textbf{b} or \textbf{t}-\textbf{c}-\textbf{d}). To obtain high-quality two-photon interference visibilities, we employed step motors to adjust the path length in order to eliminate the temporal distinguishabilities between photons 1 and 2 in the BSM. The other two Bell states ($\ket{\Phi^-}_{12}$ and $\ket{\Phi^+}_{12}$) could not be identified separately and were not further considered within the teleportation protocol. The resulting 50\% efficiency of the BSM is the optimum when using linear optical elements only~\cite{Calsamiglia2001}. Note that such a BSM yields a higher teleportation fidelity compared to the simpler interferometric implementation~\cite{Bouwmeester1997}, because it reduces noise contribution from higher-order emissions.

While the BSM was being performed at Alice, photon 3 was guided to a 7~cm aperture transmitter telescope through a 100~m long single-mode fiber and then sent via a 143~km free-space quantum channel over to Bob in Tenerife. There, it was collected by the 1~m aperture OGS telescope, and guided through its Coud\'{e} path to Bob.

In the first stage of our experiment, we only considered the cases where Alice detected $\ket{\Psi^-}_{12}$ in the BSM, which results in photon 3 being already in the state of the input photon, $|\phi\rangle_1$, and hence Bob was required to perform an identity operation, that is, do nothing at all. We verified the success of the teleportation process by analyzing photon 3's polarization state, which was accomplished by a polarization analyzer, consisting of a quarter- and a half-wave plate and two free-space coupled Si-APDs placed in each output mode of a polarizing beam splitter.

In the second stage of our experiment, we implemented a real-time feed-forward operation. When Alice obtained $\ket{\Psi^+}_{12}$, she sent this classical information to Bob. Upon receiving this information, Bob had to apply a $\pi$ phase shift between the $\ket{H}$ and $\ket{V}$ components of photon 3 to obtain the replica of the input state $|\phi\rangle_1$.

We encoded the $\ket{\Psi^+}_{12}$ BSM result in the classical pulses of a 1064~nm laser and sent them to Bob via the classical 143~km free-space channel by using a separate transmitting aperture. In Tenerife, these classical pulses, just as the quantum signal, were also collected by the 1~m aperture of the OGS telescope. Bob then used dichroic mirrors (DM) to separate the 1064~nm laser pulses from the quantum signal (photon 3), and detected the classical pulses with a photodetector (PD). A discriminator was used to convert the output of the PD into a transistor-transistor logic (TTL) signal. Thus the TTL signal carried the encoded BSM result, which was then used to trigger an electro-optical modulator (EOM) to perform the required $\pi$ phase shift operation~\cite{Ursin2004}. Finally, Bob used an extra logic circuit to locally identify the coincidence between the counts of photon 3 and these BSM results~\cite{Scheidl2010}. The 100~m long single-mode fiber was used to optically delay photon 3 at La Palma before it traveled to Tenerife, thereby giving Bob enough time to perform the $\pi$ phase shift operation.

The relevant events on Alice's and Bob's sides were recorded with separate time-tagging units each disciplined to the global positioning system (GPS)~\cite{Ursin2007}. First, Alice in La Palma identified the 3-fold coincidence event corresponding to the $\ket{\Psi^{\pm}}_{12}$ outcome of the BSM. This was done using a coincidence logic circuit featuring two separate output signals (in TTL pulses). These TTL pulses were fed into a time-tagging unit which recorded the exact time and the BSM result into a binary file. Similarly, Bob fed the signals of both detectors (stage 1, without feed-forward) or the coincidence between these signals and the BSM results sent via 1064~nm laser pulses (stage 2, with feed-forward) into his time-tagging unit. After a measurement run was completed, both time-tagged data files were compared by cross-correlation, and the detection events associated with simultaneous detection of four photons as originating from the same pump pulse were identified.

\section{Significant experimental challenges in real life}

The real-life long-distance environment implied a number of challenges for the present teleportation experiment. They resulted most significantly in the necessity to cope with an extremely low signal-to-noise ratio when using standard techniques, indeed too low for performing a successful experiment. To enhance it to a level making the experiment possible at all, we employed for the first time a combination of various cutting-edge techniques.

To increase the signal, a frequency-uncorrelated polarization-entangled photon pair source was developed providing high-rate high-quality entangled photons. The source of the input photon was optimized by employing a collinear configuration for the heralded single-photon source. To reduce the noise, ultra-low-noise single-photon detectors at Bob were employed and an entanglement-assisted technique was implemented to synchronize Alice's and Bob's clocks.

Furthermore, the severe environmental conditions imposed demanding requirements onto the whole long-distance teleportation setup: link attenuation fluctuations due to atmospheric turbulence, induced by the harsh meteorological situations including rapid temperature change, sand storms, rain, fog, strong wind and even snow. These severe conditions delayed our experimental realizations of quantum teleportation for nearly one year, as due to the exceptionally bad weather conditions, the experiment could not be carried out during our residency on the Canary Islands from May to July in 2011. We note that many of these conditions would not hamper a future satellite experiment as severely, because when communicating upwards to a satellite, much less atmosphere has to be passed by the optical signals. We will now describe the techniques employed in detail.

To eliminate the frequency correlation in the femtosecond-laser-pumped entangled photon source, we employed a spectral compensation scheme based on an interferometric Bell-state synthesizer~\cite{Kim2003,Poh2009}, as shown in Fig.\ \ref{figSetup}. This scheme allowed us to avoid the usage of narrow-band interference filters, and hence to increase the count rate while maintaining the quality of the entanglement. In our experiment, we used an 8~nm interference filter for photon 3, and 3~nm for photon 2. With the 8~nm interference filter, we almost doubled the count rate of the entangled photon source without losing significant quality of polarization entanglement. The Bell-state synthesizer ensured every photon 2 (photon 3) to be an extraordinary (ordinary) photon. In the BSM, it is important to interfere two extraordinary photons on the FBS to obtain high count rate and good visibility~\cite{Yao2012}. Therefore, we interfered photon 2 with photon 1 (the extraordinary photon generated from BBO$_2$).

The advantage of using the collinear configuration for the HSP source is that it does not require the use of polarization filters to prepare the teleportee photon 1 in a well-defined polarization state, as would be necessary in the non-collinear configuration~\cite{Bouwmeester1997}. Hence, with the collinear configuration, the rate at which the teleportee photons could be obtained was enhanced by a factor of two. The resulting two-fold coincidence rates of EPR source and HSP source were about 140~kHz and 180~kHz, respectively.

In order to reduce the background counts on Bob's side, we employed two ultra-low dark count rate free-space coupled Si-APDs (PerkinElmer C30902SH) with 500 $\textrm{$\mu$}$m active area diameter, which were passively quenched. They were biased at 11 V over-voltage and thermoelectrically cooled to $-65$ $^{\circ}$C to reduce their intrinsic dark count rate to about 15~Hz per detector in an enclosed environment~\cite{Kim2011}. Bob's measured background count rate during the experiment was about 100~Hz per detector, mainly because of stray light.

Establishing a precise common time basis between the two islands allowed us to narrow the coincidence window size and obtain a better signal-to-noise ratio. To achieve this, first a coarse time base was provided by pre-synchronization of the local clocks with GPS. However, due to the drift of the GPS clocks (about $\pm$5~ns in 990~s integration time) we would have been forced to use a large coincidence time window ($\geq$ 10~ns) and hence would have increased the noise on Bob's side if no further synchronization would have been implemented. By employing entanglement-assisted clock synchronization using two-fold coincidence counts (between detectors a and e, Fig.\ \ref{figSetup}), we were able to minimize this noise contribution. This procedure was performed every 45--180 seconds and allowed to precisely synchronize the individual time bases at Alice and Bob~\cite{Scheidl2009}. The combined timing resolution of the total detection module (including electronic jitter of the Si-APDs, resolution of the time-tagging units and time stability of the individual time bases) was approximately 1~ns. In the data analysis, we used 3~ns as the coincidence window.

To maintain the optical free-space communication links, i.e., the quantum and classical channels, under turbulent atmospherical conditions, a bidirectional closed-loop tracking system was implemented. The transmitter telescope platform was actively locked to a 532~nm beacon laser diode pointing from Tenerife to La Palma. Similarly, the OGS receiver telescope adjusted its pointing direction based on a 532~nm beacon laser attached to the transmitter telescope. For details on the link establishment and the bidirectional closed-loop tracking system, see refs [13,16].


\section{Experimental results}
\begin{figure}
\includegraphics[width=0.85\textwidth]{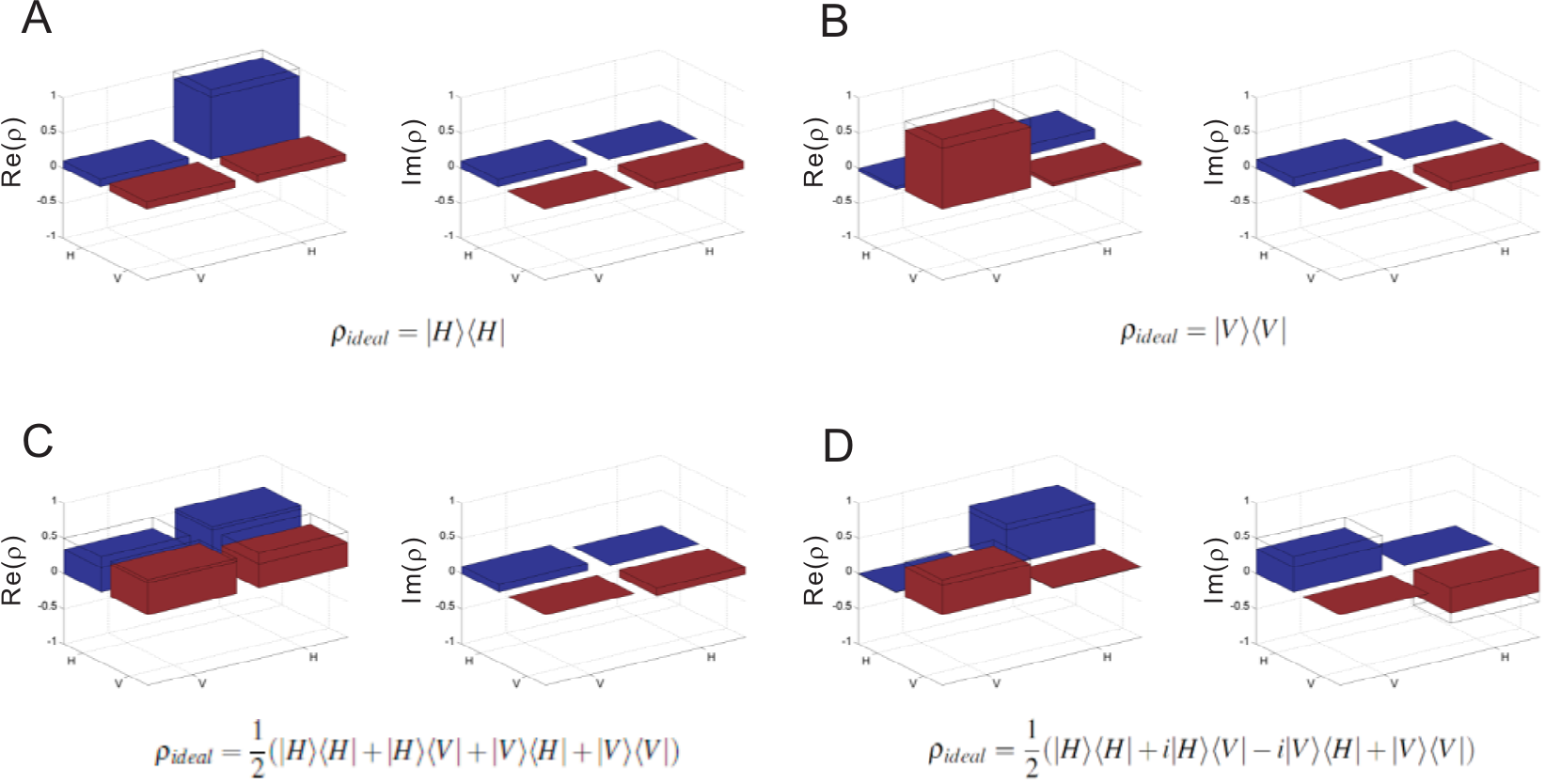}
\caption{\label{figStateTomo}State tomography results of the four quantum states without feed-forward over the 143~km free-space channel with the BSM outcome of $\ket{\Psi^-}_{12}$. The bar graphs show the reconstructed density matrices $\rho$ for the four states teleported from Alice (La Palma) to Bob (Tenerife) over the 143 km free-space channel. The wire grids indicate the expected values for the ideal cases. The data shown comprise a total of 605 four-fold coincidence counts in about 6.5 hours. The uncertainties in state fidelities extracted from these density matrices are calculated using a Monte Carlo routine assuming Poissonian errors.}
\end{figure}
\begin{figure}
\includegraphics[width=0.75\textwidth]{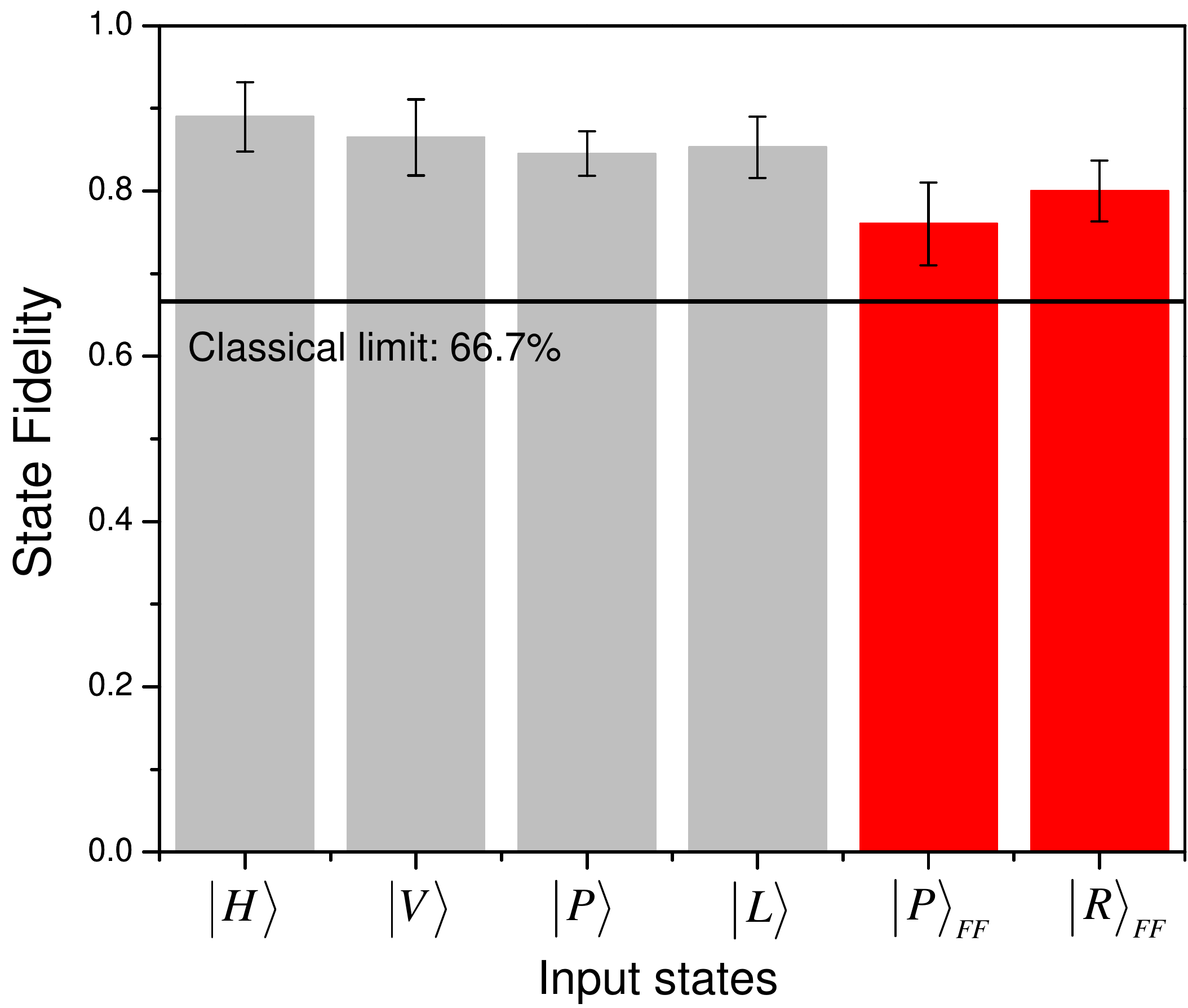}
\caption{\label{figStateFed}Summary of the state fidelity results for the teleported quantum states with and without feed-forward. Grey bars show the results obtained in quantum teleportation without feed-forward. Bob was informed via an internet connection when Alice's BSM outcome was $\ket{\Psi^-}_{12}$, i.e., those cases where he is not required to apply any operation to his photon. The uncertainties of these state fidelities are the same as presented in the caption of Fig.\ \ref{figStateTomo}. Red bars stand for the results obtained in quantum teleportation with feed-forward (FF). Bob was informed via the free-space classical link of Alice's BSM outcomes of $\ket{\Psi^+}_{12}$. The uncertainties of these state fidelities are derived from Poissonian statistics. All observed fidelities significantly exceed the classical average fidelity limit of 2/3.}
\end{figure}

First we present the results without feed-forward, where we only considered the BSM outcome $\ket{\Psi^-}_{12}$. We teleported a set of four input states $\ket{\phi_{1}}\in$
\{ $\ket{H}$, $\ket{V}$, $\ket{P}=(\ket{H}+\ket{V})/\sqrt{2}$, $\ket{L}=(\ket{H}-i\ket{V})/\sqrt{2}$ \}. We performed tomographic measurements in three consecutive nights, thereby accumulating data over 6.5 hours. Fig.\ \ref{figStateTomo} shows the state tomography results of quantum teleportation. The measured density matrix $\rho$ for each of these teleported states was reconstructed from the experimentally obtained data using the maximum-likelihood technique~\cite{White1999}. The
fidelity of the teleported state is defined as the overlap of the ideal teleported state $\ket{\phi_{ideal}}$ with the measured density matrix: $f=\left\langle\phi_{ideal}\vert\rho\vert\phi_{ideal}\right\rangle$. For this set
of states, the teleported state fidelities are measured to be $f = {0.890(42), 0.865(46), 0.845(27), 0.852(37)}$, yielding an average $\bar{f} =0.863(38)$. During these measurements the link attenuation varied from $28.1$~dB to $39.0$~dB, which was mainly caused by rapid temperature change and strong wind. Despite such high loss in the quantum free-space channel, the classical average fidelity limit~\cite{Bennett1993} of 2/3 was clearly surpassed by our observed fidelities, as shown in Fig.\ \ref{figStateFed}. Therefore, we explicitly demonstrated quantum teleportation over the 143~km free-space channel.

\begin{figure}
\includegraphics[width=0.85\textwidth]{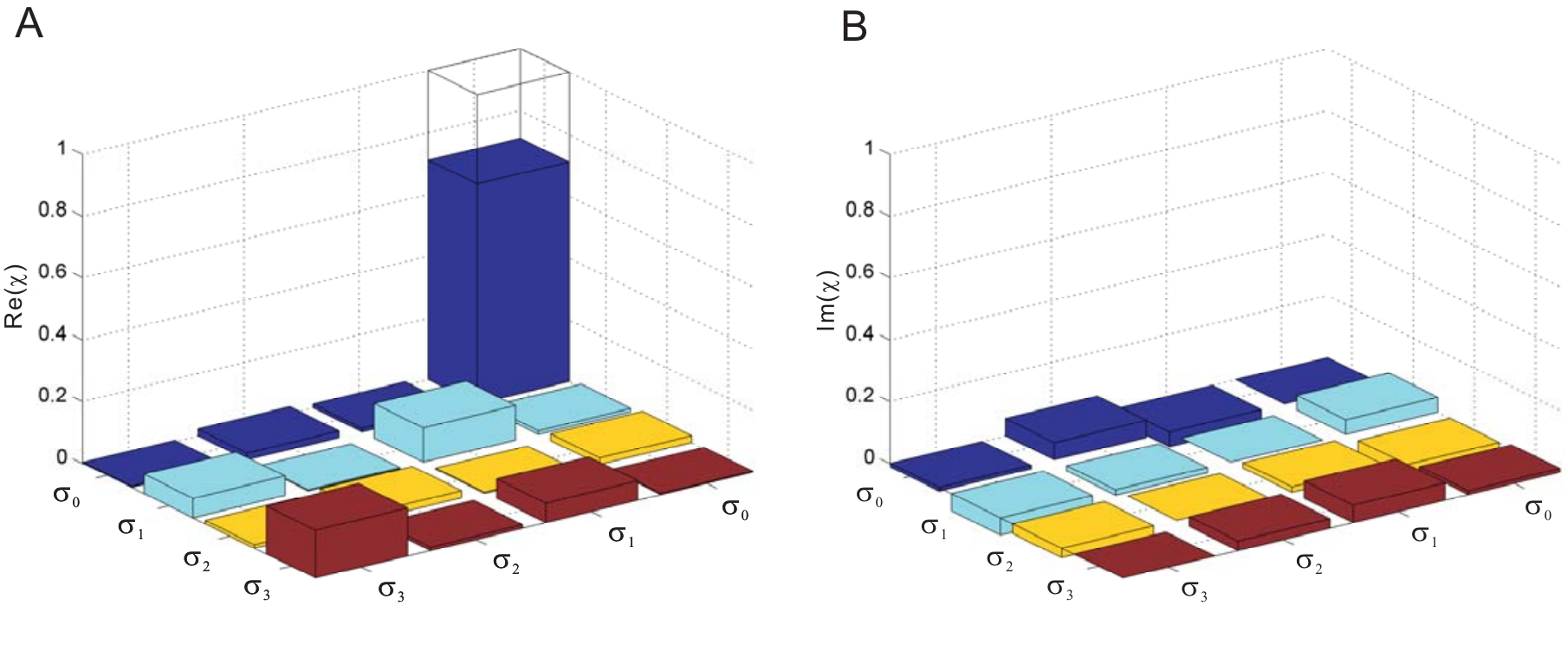}
\caption{\label{figProcesstomo}Quantum process tomography of quantum teleportation without feed-forward. The real (Re$(\chi_{lk})$) and imaginary (Im$(\chi_{lk})$) values of the components of the reconstructed quantum process matrix are shown in (A) and (B), with $l$, $k$ = 0, 1, 2, and 3. The results of the state tomography of the four teleported states, $\ket{H}$, $\ket{V}$, $\ket{P}$, $\ket{L}$, are employed to reconstruct the process matrix of quantum state teleportation. The operators $\sigma_{i}$ are the identity (\textit{i} = 0) and the \textit{x}-, \textit{y}-, and \textit{z}-Pauli matrices (\textit{i} = 1, 2, 3). For the ideal case, the only nonzero component of the process matrix of quantum teleportation, $\chi_{ideal}$, is $(\chi_{ideal})_{00}=1$, which is indicated by the wire grids. Our experiment clearly confirmed that $\chi_{00}$ (identity operation) is indeed the dominant component.}
\end{figure}

The reconstructed density matrices of the teleported quantum states allow us to fully characterize the teleportation procedure by quantum process tomography. The four input states ($\rho_{ideal} = \ket{\phi_{ideal}} \bra{\phi_{ideal}}=\ket{H}\bra{H}, \ket{V}\bra{V}, \ket{P}\bra{P}, \ket{L}\bra{L}$) are transferred to the corresponding (reconstructed) output states $\rho$. We can completely describe the effect of teleportation on the input states $\rho_{in}$ by determining the process matrix $\chi$, defined by $\rho = \sum_{l,k = 0}^3 \chi_{lk} \sigma_l \rho_{in} \sigma_k$, where the $\sigma_{i}$ are the Pauli matrices with $\sigma_{0}$ the identity operator. The process matrix $\chi$ can be computed analytically from these four equations for the four different input and output states~\cite{NielsenChuang2000}. The ideal process matrix of quantum teleportation $\chi_{ideal}$ has only one nonzero component, $\left( \chi_{ideal} \right)_{00} = 1$, meaning the input state is teleported without any reduction in quality. Fig.~\ref{figProcesstomo}A and B show the real and imaginary components of $\chi$ for quantum teleportation based on our experimental results. The process fidelity of our experiment was $f_{process} = \textrm{tr} \left( \chi_{ideal} \chi \right) = 0.710(42)$. This clearly confirmed the quantum nature of our teleportation experiment as it is 5~standard deviations above the maximum process fidelity of 0.5, which is the limit one can reach with a classical strategy where Alice and Bob do not share any entanglement as a resource.

In the second stage of the experiment, we realized quantum teleportation including real-time feed-forward of the BSM result over the 143~km classical channel. We set $\ket{P}$ and $\ket{R}$ states ($\ket{R}=(\ket{H}+i\ket{V})/\sqrt{2}$) as input states for which the required $\pi$ phase shift between the $\ket{H}$ and the $\ket{V}$ components of photon 3 resulted in a 90$^{\circ}$ polarization rotation. However for the $\ket{H}$ or
$\ket{V}$ input state, feed-forward is irrelevant because a $\pi$ phase shift would only result in a non-detectable global phase shift. Thus the quality of teleportation of these states is already confirmed by our first-stage experiment. Realizing teleportation for the states $\ket{P}$ and $\ket{R}$ fully confirms the generality of the procedure as these states belong to different mutually unbiased bases. In Tenerife, we analyzed photon 3 in the eigenbasis of the input state, i.e., the $\ket{P}/\ket{M}$ ($\ket{R}/\ket{L}$) basis when the input state was $\ket{P}$ ($\ket{R}$). Note that $\ket{M}=(\ket{H}-
\ket{V})/\sqrt{2}$. The resultant fidelities of the teleported states are 0.760(50) and 0.800(37) for $\ket{P}$ and $\ket{R}$, respectively (red bars in Fig.\ \ref{figStateFed}). Both results are clearly above the classical fidelity bound. Note that in our experiment, the efficiency of the classical link was about 21.3\%, which was mainly due to amplitude fluctuations caused by atmospherical turbulence.

Our results show that quantum teleportation using active feed-forward can be achieved even over these large free-space distances under real outdoor conditions, which is of utmost importance in future quantum information applications. We note that from a conceptual perspective real-time feed-forward is part of the original teleportation proposal~\cite{Bennett1993}.

It might be noted that photon 3 was still in the 100~m fiber at the JKT telescope when the BSM took place. We suggest that this is not a limitation because the teleportation protocol itself is only completed after Bob has received the classical information about Alice's BSM outcome and after he has applied the corresponding unitary transformation. Only then a replica of the input state is reproduced at his distant location and the quantum teleportation protocol is thus successfully finished. Alice's BSM itself does not conclude the teleportation process.

\section{Concluding comments}

Using the real-time feed-forward operation, we unambiguously experimentally demonstrated quantum teleportation from La Palma to Tenerife over a 143 km free-space channel. Ultimately, the advantage of long-distance teleportation compared to just sending the quantum state itself may lie in the following future applications: If Alice and Bob can stockpile their EPR states beforehand (with the help of quantum memories), teleportation is advantageous, if the quantum channel is of low quality or if Bob's location is unknown to Alice. This is because Alice can broadcast the classical information with high quality and to different locations~\cite{Bennett1993}. Also the quantum repeater, which is of high importance for large-scale quantum networks, is based on teleportation in the form of entanglement swapping~\cite{Zukowski1993}.

Our work proves the feasibility of both ground-based and satellite-based free-space quantum teleportation. Our quantum teleportation setup was able to achieve coincidence production rates and fidelities to cope with the optical link attenuation resulting from various experimental and technical challenges, which will arise in a quantum transmission between a ground-based transmitter and a low-earth-orbiting (LEO) satellite receiver~\cite{Aspelmeyer2003b}. Actually, some demands in our experiment were even more challenging since in satellite communication the atmospheric distances to be overcome are certainly shorter than the distance between Tenerife and La Palma. Therefore, our experiment represents a crucial step toward future quantum networks in space, which require space to ground quantum communication. The technology implemented in our experiment thus certainly reached the required maturity both for satellite and for long-distance ground communication. We expect that many of the features implemented here will be key blocks for a new area of fascinating experiments.

\section*{Acknowledgements}
The authors thank the staff of IAC: F. Sanchez-Martinez, A. Alonso, C. Warden, M. Serra, J. Carlos and the staff of ING: M. Balcells, C. Benn, J. Rey, O. Vaduvescu, A. Chopping, D. Gonz\'{a}ez, S. Rodr\'{\i}guez, M. Abreu, L. Gonz\'{a}ez; J. Kuusela, E. Wille, Z. Sodnik, and J. Perdigues of the OGS and ESA. X.S.M., T.J., R.U. and A.Z. thank S. Ramelow for helpful discussion, P. Kolenderski for helpful discussions on the SPDC source with the Bell-state synthesizer, S. Zotter for help during the early stages of the experiment and Prof. Steinacker for the expert meteorological advice. J.K. acknowledges support by the EU project QUEVADIS. E.A. and V.M. thank C.~Kurtsiefer and Y.-S.~Kim for detector electronics design, J.~Skaar, Research Council of Norway (grant No.\ 180439/V30) and Industry Canada for support. This work was made possible by grants from the European Space Agency (Contract 4000104180/11/NL/AF), the Austrian Science Foundation (FWF) under projects SFB F4008 and CoQuS, and the FFG for the QTS project (No. 828316) within the ASAP 7 program. We also acknowledge support by the European Commission, grant Q-ESSENCE (No. 248095) and the John Templeton Foundation.


\end{document}